\documentclass[12pt,preprint]{emulateapj}

\def\be{\begin{equation}}
\def\ee{\end{equation}}
\def\bea{\begin{eqnarray}}
\def\eea{\end{eqnarray}}
\usepackage{graphicx}
\usepackage{amsmath}
\usepackage[colorlinks,
            linkcolor=blue,
            anchorcolor=blue,
            citecolor=blue]{hyperref}

\usepackage{color,txfonts}
\usepackage{threeparttable}
\begin{document}

\title{Detection of GeV $\gamma$-ray emission in the direction of HESS J1731-347 with Fermi-LAT}

\author{Xiao-Lei Guo$^{1,2}$, Yu-Liang Xin$^{1,3}$, Neng-Hui Liao$^{1,2}$, Qiang Yuan$^{1,2}$, Wei-Hong Gao$^{4,5}$, Yi-Zhong Fan$^{1,2}$}


\affil{$^1$Key laboratory of Dark Matter and Space Astronomy, Purple Mountain Observatory, Chinese Academy of Sciences, Nanjing 210008, China}

\affil{$^2$School of Astronomy and Space Science, University of Science and Technology of China, Hefei 230026, Anhui, China}

\affil{$^3$University of Chinese Academy of Sciences, Yuquan Road 19, Beijing, 100049, China}

\affil{$^4$Department of Physics and Institute of Theoretical Physics, Nanjing Normal University, Nanjing 210046, China}

\affil{$^5$INAF-Osservatorio Astronomico di Brera, via E. Bianchi 46, 23807 Merate, Italy\\
ylxin@pmo.ac.cn (YLX); yuanq@pmo.ac.cn (QY); gaoweihong@njnu.edu.cn (WHG)
}

\begin{abstract}
We report the detection of GeV $\gamma$-ray emission from supernova remnant 
HESS J1731-347 using 9 years of {\it Fermi} Large Area Telescope data. 
We find a slightly extended GeV source in the direction of HESS J1731-347. 
The spectrum above 1 GeV can be fitted by a power-law with an index of 
$\Gamma = 1.77\pm0.14$, and the GeV spectrum connects smoothly with the 
TeV spectrum of HESS J1731-347. 
Either a hadronic-leptonic or a pure leptonic model can fit the 
multi-wavelength spectral energy distribution of the source. However,
the hard GeV $\gamma$-ray spectrum is more naturally produced in a 
leptonic (inverse Compton scattering) scenario, under the framework
of diffusive shock acceleration. We also searched for the GeV $\gamma$-ray 
emission from the nearby TeV source HESS J1729-345. No significant GeV 
$\gamma$-ray emission is found, and upper limits are derived.

\end{abstract}


\setlength{\parindent}{.25in}

\section{Introduction}
It is widely believed that supernova remnants (SNRs) are the main 
accelerators of Galactic cosmic rays (CRs) with energies up to the knee. 
This is supported by the non-thermal X-ray emission detected in many
SNRs, which indicates the acceleration of electrons to hundreds of TeV
energies \citep[e.g.,][]{Koyama1995}.
The GeV and/or TeV $\gamma$-rays have also been detected in some SNRs, 
for example, RCW 86 \citep{Aharonian2009,Yuan2014};
Cas A \citep{Albert2007a, Abdo2010a}; 
CTB 37B \citep{Aharonian2008a, Xin2016}; 
Puppis A \citep{Hewitt2012, Xin2017}; 
IC 443 \citep{Albert2007b, Acciari2009, Ackermann2013}; 
W44 \citep{Abdo2010b, Ackermann2013}.
Gamma-rays can be produced by the decay of neutral pions due to the 
inelastic $pp$ collisions (the hadronic process), the Inverse Compton 
Scattering (ICS) or bremsstrahlung process of relativistic electrons 
(the leptonic process). For some SNRs interacting with dense molecular
clouds, the evidence for acceleration of nuclei has been suggested
by GeV/TeV $\gamma$-ray observations \citep{Li2010,Li2012,Ackermann2013}.

HESS J1731-347 (G353.6-0.7) was first observed as an unidentified 
very-high-energy (VHE; $>$100 GeV) $\gamma$-ray source by the 
High Energy Stereoscopic System (HESS) \citep{Aharonian2008b}. 
\citet{Tian2008} discovered the radio and X-ray counterparts of 
HESS J1731-347 and identified it as a shell-type SNR. 
\citet{Abramowski2011} carried out an additional $\gamma$-ray 
observation with HESS and detected its shell-type morphology. 
Together with RX J1713.7-3946 \citep{Aharonian2004,Aharonian2006,
Aharonian2007a}, RX J0852.0-4622 \citep{Aharonian2005, Aharonian2007b}, 
RCW 86 \citep{Aharonian2009,Abramowski2016} and SN 1006 \citep{Acero2010}, 
HESS J1731-347 becomes
one of five firmly identified TeV shell-type SNRs \citep{Rieger2013}.
 


The distance of HESS J1731-347 is under debated. \citet{Tian2008} argued 
that HESS J1731-347 locates at $\sim$3.2 kpc if it is associated with the 
nearby HII region G353.42-0.37. By comparing the absorption column density 
derived from the X-ray observation and that obtained from $^{12}$CO and 
HI observations, \citet{Abramowski2011} set 3.2 kpc as a lower limit of
its distance, which is reinforced by \citet{Doroshenko2017}. 
In addition, \citet{Fukuda2014} suggested that HESS J1731-347 is correlated 
with the interstellar proton cavity at a velocity range from $-90$ km 
s$^{-1}$ to $-75$ km s$^{-1}$, indicating a distance of 5.2$-$6.1 kpc. 
However, no significant emission from dense molecular gas traced by CS(1-0) line
coincides with HESS J1731-347 at that distance \citep{Maxted2017}.
Due to uncertainties of the distance and other parameters, 
the age of HESS J1731-347 is estimated to be in a wide range of $2-27$ kyr 
\citep{Tian2008,Abramowski2011,Fukuda2014,Acero2015b}.

HESS J1731-347 and its sub-regions were detected in the X-ray band by 
{\it ROSAT}, {\it XMM-Newton} and {\it Suzaku}, with an X-ray morphology 
consistent with the radio shell \citep{Tian2008,Tian2010,
Abramowski2011,Bamba2012,Doroshenko2017}. 
The X-ray emission from the complete SNR and its sub-regions are found 
to be non-thermal. 
The X-ray spectral index is 2.66 for the entire SNR \citep{Doroshenko2017}, 
and is somewhat harder ($\Gamma=2.28$) for the north-east region 
\citep{Abramowski2011}.
A compact object, XMMS J173203-344518, located near the geometrical center 
of the remnant, was detected by {\it XMM-Newton}, which was considered to 
be the central compact object (CCO) associated with HESS J1731-347 
\citep{Tian2010,Halpern2010,Abramowski2011}.

\citet{Yang2014} and \citet{Acero2015b} searched for the GeV $\gamma$-ray 
emission from HESS J1731-347 with the {\it Fermi} Large Area Telescope
\citep[{\it Fermi}-LAT;][]{Atwood2009} data. No significant signal was detected and only the 
upper limits were given.
Furthermore, no candidate source in the third {\it Fermi}-LAT source catalog 
\citep[3FGL;][]{Acero2015a} is found to be associated with HESS J1731-347.

In this paper, we revisit the GeV $\gamma$-ray emission in the direction 
of HESS J1731-347, with 9 year Pass 8 data recorded by {\it Fermi}-LAT. 
A statistically significant excess which is positionally consistent with
HESS J1731-347 is found. In Section 2, we present the data analysis and 
results, including the spatial and spectral analysis. Based on the 
multi-wavelength observations of HESS J1731-347, we model the non-thermal 
radiation of it in Section 3. The conclusion of this work is presented in 
Section 4.

\section{Data Analysis}

\subsection{Data Reduction}

\begin{figure*}[!htb]
\includegraphics[width=3.5in]{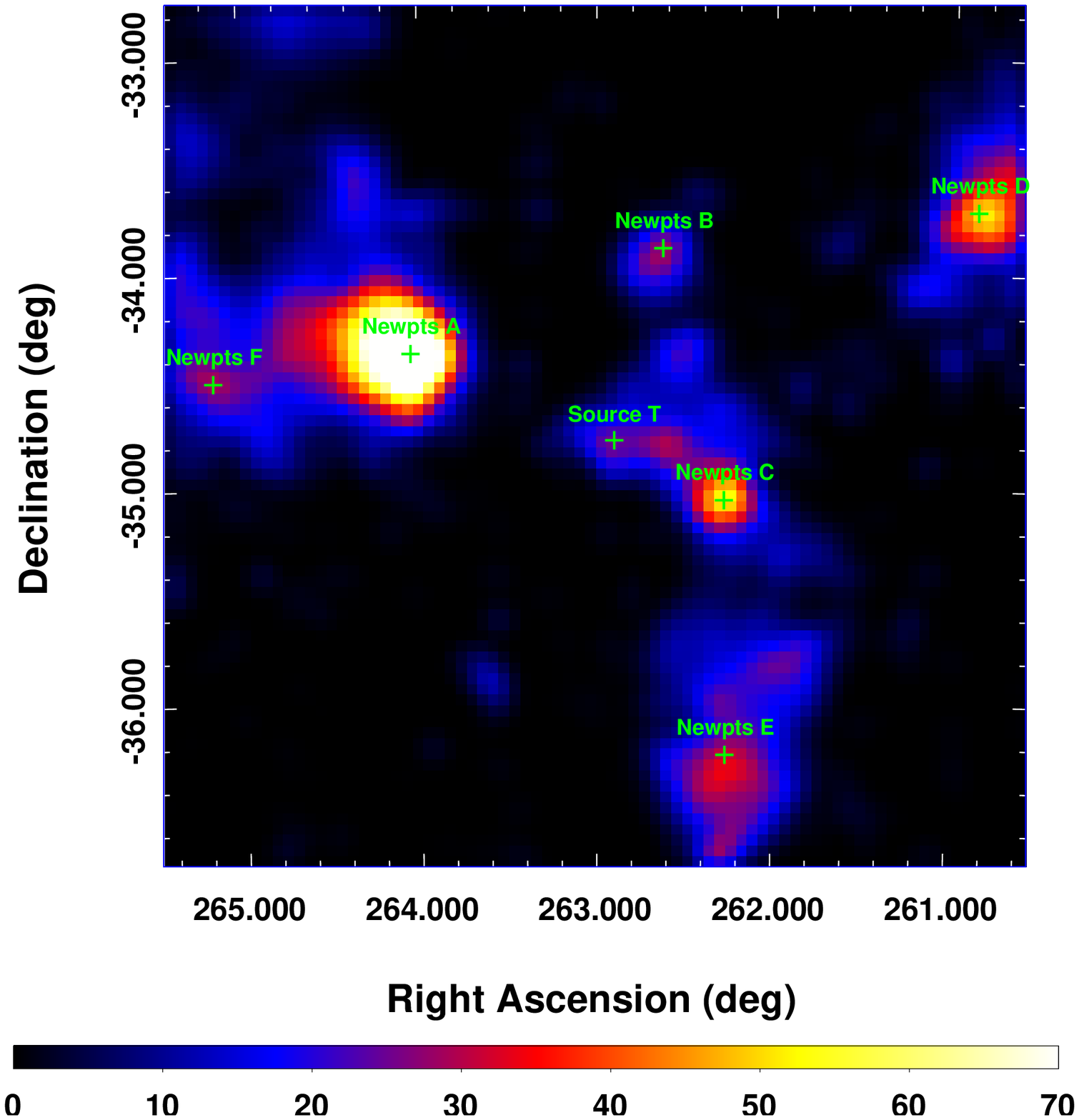}
\includegraphics[width=3.5in]{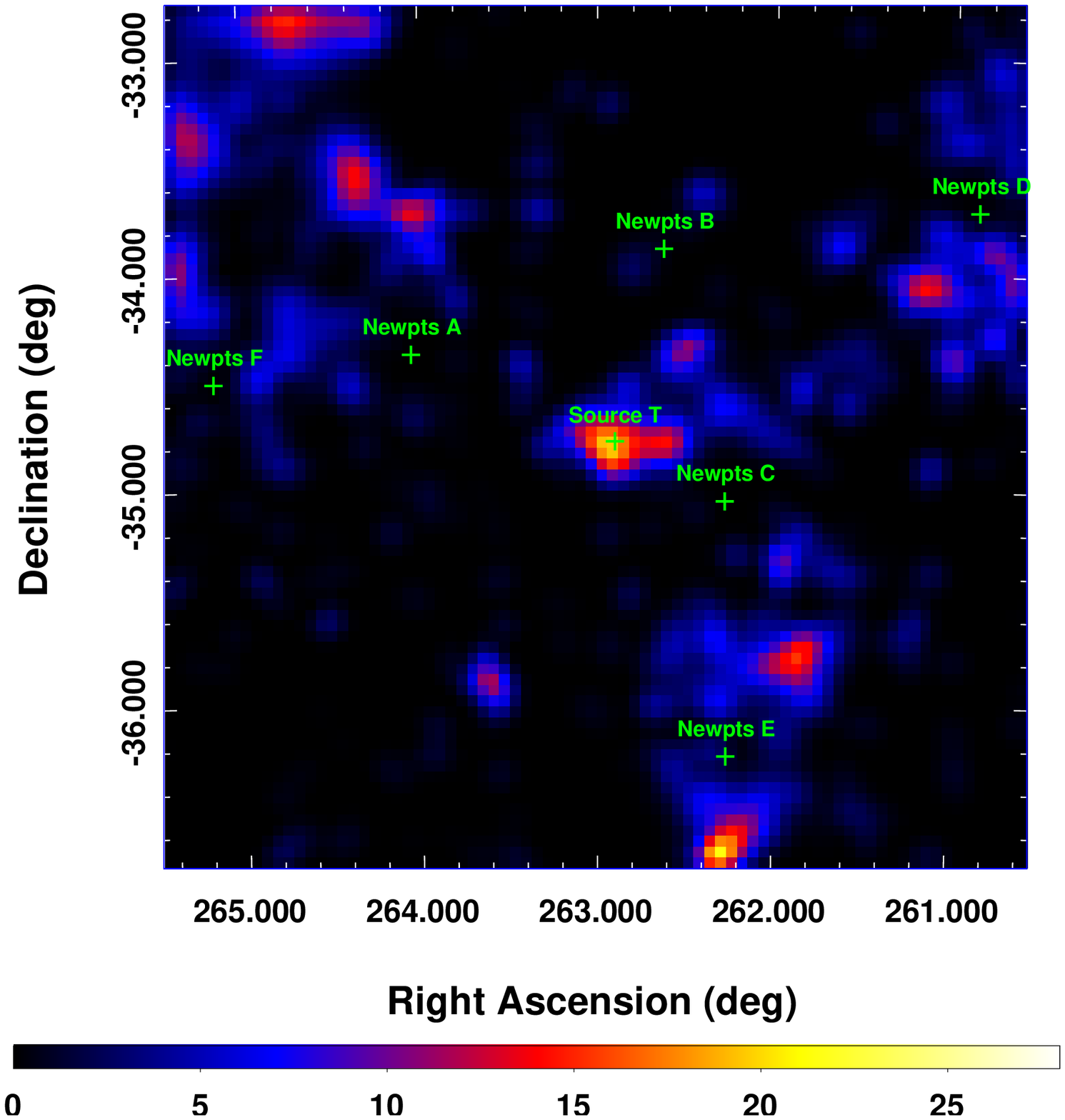}
\caption{TS maps of a $4^{\circ}\times4^{\circ}$ region centered on the 
position of HESS J1731-347, for photons above 1 GeV. The left panel is 
the TS map with 3FGL sources and the diffuse backgrounds subtracted,
and the right panel is the TS map with the six additional new sources 
A to F (green crosses) subtracted. 
All maps are smoothed with a Gaussian kernel with $\sigma$ = $0.1^\circ$.}
\label{fig:tsmap}
\end{figure*}


We select the latest Pass 8 version of the {\it Fermi}-LAT data with 
``Source'' event class (evclass=128 \& evtype=3), recorded from August 
4, 2008 (Mission Elapsed Time 239557418) to August 4, 2017 (Mission 
Elapsed Time 523497605). 
The region of interest (ROI) is chosen to be a $14^\circ \times 14^\circ$ 
box centered at HESS J1731-347. In order to have a good angular resolution, 
we adopt the events with energies between 1 GeV and 300 GeV in this analysis.
In addition, the events whose zenith angles are larger than $90^\circ$ are 
excluded to reduce the contamination from the Earth Limb. The data are 
analyzed with the {\it Fermi}-LAT Science Tools 
v10r0p5\footnote{http://fermi.gsfc.nasa.gov/ssc/data/analysis/software/},
and the standard binned likelihood analysis method {\tt gtlike}.
The diffuse backgrounds used are {\tt gll\_iem\_v06.fits} and 
{\tt iso\_P8R2\_SOURCE\_V6\_v06.txt}, which can be found from the 
Fermi Science Support Center\footnote{http://fermi.gsfc.nasa.gov/ssc/data/access/lat/BackgroundModels.html}.
All sources listed in the 3FGL and the two diffuse backgrounds are 
included in the model.
During the fitting procedure, the spectral parameters and the normalizations 
of sources within $5^\circ$ around HESS J1731-347, together with the 
normalizations of the two diffuse backgrounds, are left free.

\subsection{Results}
We create a $4^\circ \times 4^\circ$ TS (test statistic, which is
essentially the logarithmic likelihood ratio between different models) 
map centered at HESS J1731-347 by slice the center of the box along 
each axis, after subtracting the 3FGL sources and the diffuse backgrounds.
There are still
excesses in this TS map, as marked out by green crosses. At the center of
the TS map, a weak excess (labelled as Source T) is found to be spatially 
coincident with HESS J1731-347. 
It is noted that Newpts C was also detected in \citet{Yang2014} with a TS value of about 20.
We add Source T and the other six new 
sources, from A to F, in the model as additional point sources with 
power-law (PL) spectra, and re-do the likelihood fitting. The positions 
of these new sources are optimized by the {\tt gtfindsrc} tool. 
Best-fitting results of their coordinates and TS values are listed in 
Table \ref{table:newpts}. The TS value of Source T is about 25.9, and 
its best-fitting position is R.A.$=262.902^\circ$, Dec.$=-34.775^\circ$ 
with 1$\sigma$ error circle of $0.022^\circ$. The residual TS map after
subtracting the additional sources A to F is shown in the right panel
Figure \ref{fig:tsmap}.

\begin{figure}[!htb]
\centering
\includegraphics[width=3.5in]{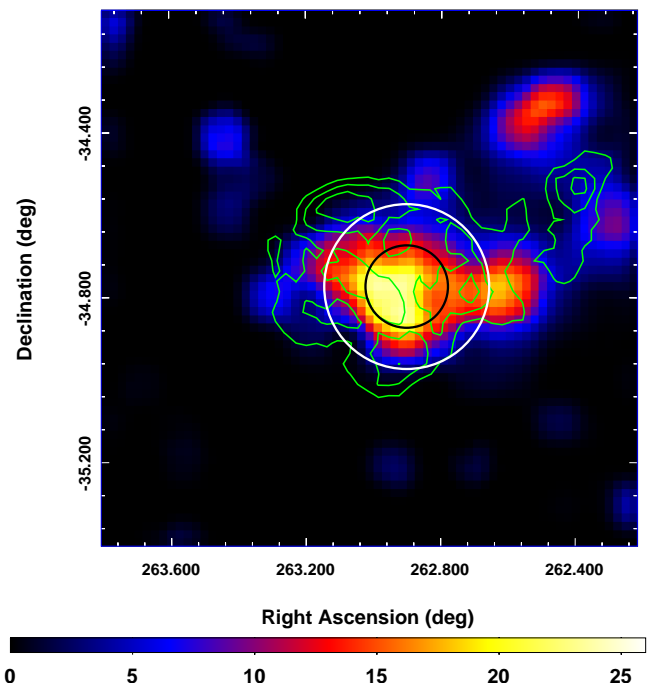}
\caption{Zoom-in of the right panel of Figure \ref{fig:tsmap}, for a region 
of $1.3^\circ \times 1.3^\circ$ centered at HESS J1731-347, overlaid with the
contours of the VHE image by HESS \citep{Abramowski2011}. The green contours
to the west show the VHE image of HESS J1729-345. The radius of Gaussian 
smooth kernel is $\sigma = 0.04^\circ$. 
The black and white circles mark the radii 
of $0.1^\circ$ and $0.2^\circ$ uniform disk, respectively.}
\label{fig:tsmap-small}
\end{figure}

Figure \ref{fig:tsmap-small} gives a $1.3^\circ \times 1.3^\circ$ zoom-in of 
Figure \ref{fig:tsmap}b in order to better show the spatial distribution 
of the target source T and its relationship with HESS J1731-347. 
The GeV $\gamma$-ray emission overlaps with part of the VHE emission 
region shown by the contours \citep{Abramowski2011}. However, the GeV 
TS map does not fully overlap with the TeV image, which is possibly due
to the large point spread function (PSF) of Fermi-LAT and/or the 
fluctuation of the weak signal. Similar cases were also shown for 
SN 1006 \citep{Xing2016} and HESS J1534-571 \citep{Araya2017}.

\begin{table}[!htb]
\centering
\caption {Coordinates, TS values, and angular separations from the center
of HESS J1731-347 of the newly added point sources}
\begin{tabular}{ccccc}
\hline \hline
Name & R.A. & Dec. & TS  & $\Delta\theta$ \\
     & [deg]&[deg] &    & [deg]\\

\hline
Source T      & $262.902$ & $-34.775$ & $25.9$  & $0.093$ \\
Newpts A      & $264.048$ & $-34.370$ & $130.8$ & $0.936$ \\
Newpts B      & $262.629$ & $-33.882$ & $29.3$  & $0.929$ \\
Newpts C      & $262.280$ & $-35.051$ & $59.8$  & $0.670$ \\
Newpts D      & $260.867$ & $-33.704$ & $59.5$  & $2.062$ \\
Newpts E      & $262.266$ & $-36.233$ & $34.9$  & $1.598$ \\
Newpts F      & $265.161$ & $-34.500$ & $26.5$  & $1.786$ \\  
\hline
\hline
\end{tabular}
\label{table:newpts}
\end{table}

\subsubsection{Spatial Extension}

Considering that HESS J1731-347 has an extended morphology in radio,
X-ray, and TeV $\gamma$-ray bands, we carried out an extension test with 
different spatial models. We used a uniform disk centered at the best-fitting 
position of Source T with radius of $0.1^\circ$, $0.15^\circ$, $0.2^\circ$, 
and $0.25^\circ$, as well as the TeV $\gamma$-ray image of HESS J1731-347, 
as spatial templates of Source T. The TS values for different spatial 
models are listed in Table \ref{table:template}. We found that a 
$0.15^\circ$ disk template gives the highest TS value, 33.9, which
corresponds to a significance of $\sim4.7\sigma$ for five (2 for the
coordinates, 1 for the radius, and 2 for the spectrum) degrees of 
freedom (dof). For the four adopted disk templates, the TS values do
not differ much from each other. Compared with the point source hypothesis,
the data favors slightly an extended morphology. Using the TeV $\gamma$-ray
template, a TS value of 25 is found. 
These results are quite consistent with that of \citet{Condon2017}, which used
the data with different energy ranges and observation time-series.
In the following analysis, we adopt the $0.15^\circ$ disk template for Source T. 

We also try to search for $\gamma$-ray emission from the nearby TeV source 
HESS J1729-345. The TeV image of HESS J1729-345 is used as the spatial 
template. No significant GeV $\gamma$-ray emission from the direction of
HESS J1729-345 is detected. The TS value of HESS J1729-345 is about 4, 
and its flux upper limits will be derived (see the next sub-section).

\begin{table}[!htb]
\centering
\caption {TS values of Source T with different spatial models}
\begin{tabular}{cccc}
\hline \hline
Spatial Model & TS & Degrees of Freedom \\
\hline
Point Source                  & $25.9$ & $4$ \\
$0.1^\circ$ Uniform Disk      & $32.1$ & $5$  \\
$0.15^\circ$ Uniform Disk      & $33.9$ & $5$  \\
$0.2^\circ$ Uniform Disk      & $32.4$ & $5$  \\
$0.25^\circ$ Uniform Disk      & $32.4$ & $5$  \\
TeV Image                     & $25.0$ & $2$  \\
\hline
\hline
\end{tabular}
\label{table:template}
\tablecomments{The 4 dof for the point source model include 2 spatial 
and 2 spectral parameters. The uniform disk has 5 dof, 2 for coordinate, 2 for spectral parameters and 1 for radius.
For the TeV image template model, only 2 dof of the spectral parameters 
are considered.}
\end{table}

\subsubsection{Spectral Analysis}
For Source T, the global fit in the 1$-$300 GeV energy range with a $0.15^\circ$ disk template 
gives a spectral index of $\Gamma = 1.77\pm0.14$, and an integral photon flux of 
$(6.92\pm2.06)\times10^{-10}$ photon cm$^{-2}$ s$^{-1}$ with statistical errors only.
Assuming a distance of 3.2 kpc \citep{Tian2008,Nayana2017}, the 
$\gamma$-ray luminosity between 1 GeV and 300 GeV is 
$1.26\times 10^{34}\,(d/3.2\ {\rm kpc})^2$ erg~s$^{-1}$.

\begin{figure}[!htb]
\centering
\includegraphics[angle=0,scale=0.35,width=0.5\textwidth,height=0.3\textheight]{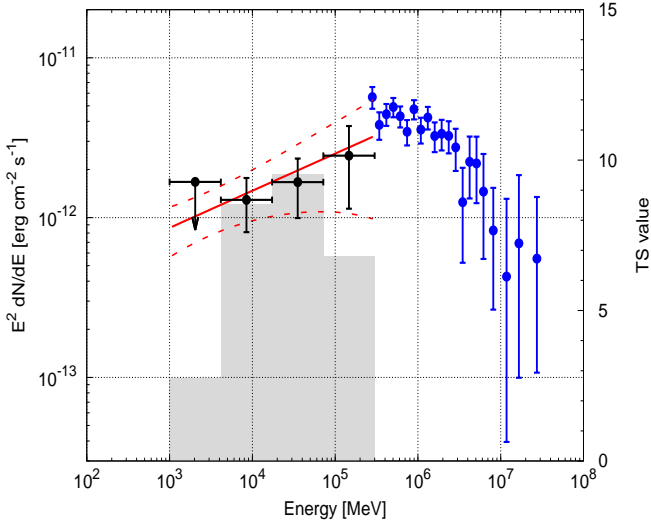}%
\hfill
\caption{The Fermi-LAT SED of Source T (black dots). The red solid and 
dashed lines show the best-fitting power-law spectrum and its 1$\sigma$ 
statistic error band. Shaded gray regions (right axis) show the TS values 
of the four energy bins. The blue dots are the HESS observations of HESS 
J1731-347 in the VHE band \citep{Abramowski2011}.}
\label{fig:sed}
\end{figure}

The data is further divided into four energy bins with equal width in the
logarithmic space to study is spectral energy distribution (SED). For each 
energy bin, we repeat the likelihood analysis, with only the normalizations 
of the sources within $5^\circ$ around Source T and the diffuse backgrounds 
in the model free. The spectral parameters of these sources are fixed to be 
the best-fitting values obtained in the global likelihood analysis. If the 
TS value of Source T is smaller than 4 in an energy bin, a 95\% confidence 
level upper limit is given. The results of the SED are shown in Figure 
\ref{fig:sed}. The GeV SED connects smoothly with the TeV 
spectrum of HESS J1731-347. The spatial coincidence and a smoothly connected 
$\gamma$-ray spectrum suggest that Source T is the GeV counterpart of 
HESS J1731-347.

\begin{figure}[!htb]
\centering
\includegraphics[angle=0,scale=0.35,width=0.5\textwidth,height=0.3\textheight]{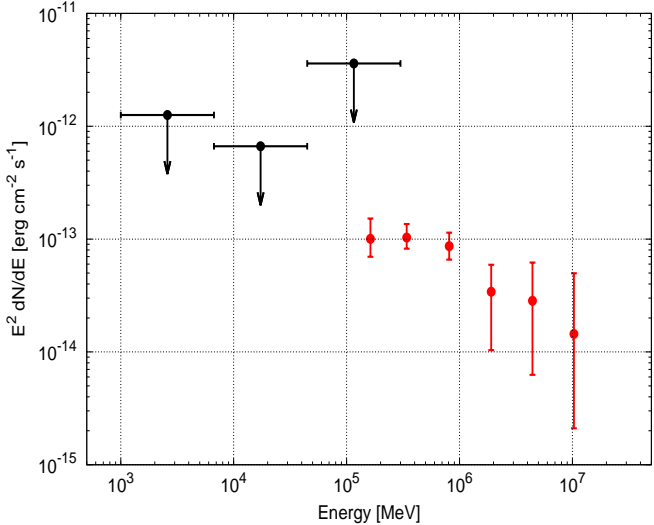}%
\hfill
\caption{Upper limits of GeV $\gamma$-ray emission from the direction
of HESS J1729-345, together with the HESS observations in the VHE band
\citep{Abramowski2011}.}
\label{fig:sed2}
\end{figure}

The significance of HESS J1729-345 is not high enough, and we derive the 
flux upper limits in energy bins of $1-6.7$, $6.7-44.8$, and $44.8-300$ GeV,
which are shown in Figure \ref{fig:sed2}. 

\section{Discussion}



The radio counterpart of HESS J131-347 was firstly identified by 
\citet{Tian2008}. The integrated flux density was derived to be $2.2 \pm 0.9$ 
Jy at 1420 MHz, through extrapolating that of one half of the remnant at 
low Galactic latitudes to the total SNR. With the Giant Metrewave Radio 
Telescope (GMRT), \citet{Nayana2017} observed the complete shell of 
HESS J131-347 at 325 MHz, and obtained an integrated flux density of 
$1.84 \pm 0.15$ Jy. 
In the following models, we use the result of \citet{Nayana2017} 
to constrain the model parameters.
The X-ray flux of the full SNR given by \citet{Doroshenko2017} is also used. 


We assume either a pure leptonic model or a hadronic-leptonic hybrid one
to fit the wide-band SED from radio to TeV $\gamma$-rays. The spectrum of 
electrons or protons is assumed to be an exponential cutoff power-law form
\begin{equation}
dN/dE \propto E^{-\alpha_{i}} {\rm exp}[-(E/E_{c,i})^{\delta}]\,, \nonumber
\end{equation}
where {\emph i} = {\emph e} or {\emph p}, $\alpha_{i}$ is the spectral 
index, $E_{c,i}$ is the cutoff energy of particles. 
and $\beta$ describes the sharpness of the cutoff. 
$\delta$ describes the sharpness of the cutoff, 
and we adopt the typical values of 0.5, 0.6 and 1.0 to constrain the parameters in the model.
The radius of the SNR is nearly $0.25^\circ$ in the radio band
\citep{Tian2008,Nayana2017}, and $0.27^\circ$ in the TeV band
\citep{Abramowski2011}. Such an angular size corresponds to a physical 
radius of about $14-15$ pc for a distance of 3.2 kpc. The gas density in
the vicinity of HESS J1731-347 is quite uncertain, due to the lack of
thermal X-ray emissions. We assume a nominal value of $n = 1.0$ cm$^{-3}$.

\begin{figure*}[!htb]
\includegraphics[width=3.5in]{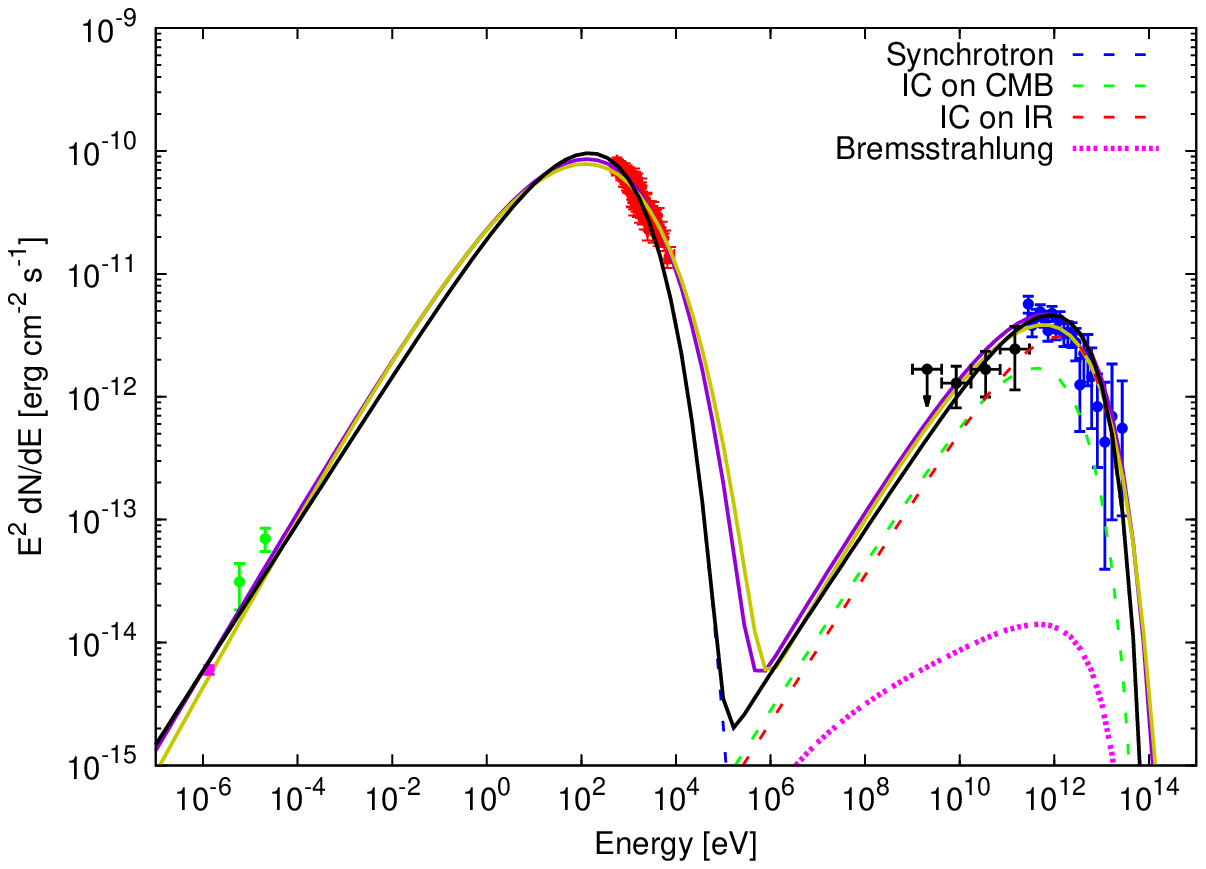}
\includegraphics[width=3.5in]{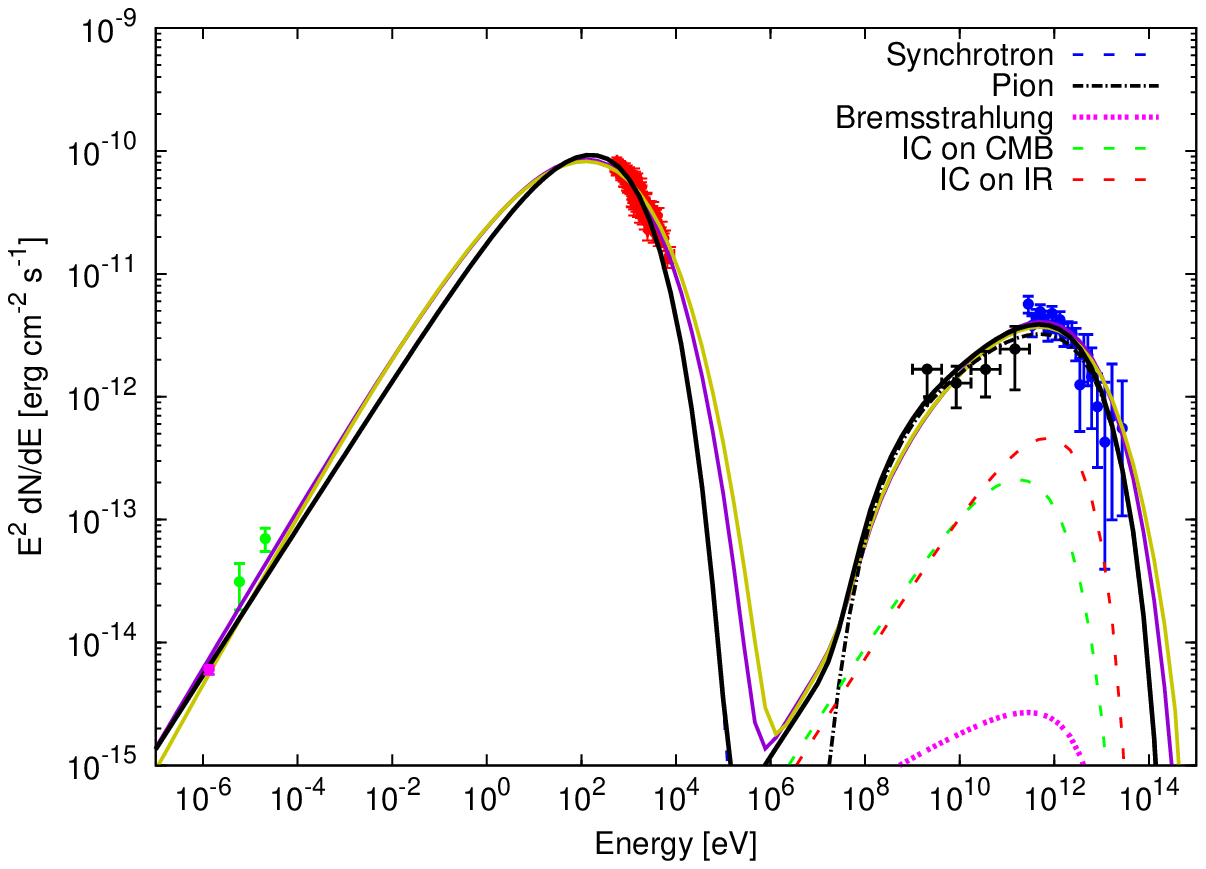}
\caption{Modeling of the multi-wavelength SED of HESS J1731-347. 
The left panel is for the leptonic model, and the right panel is for the hadronic-leptonic hybrid model. 
The models with $\delta = 0.5,~0.6,$ and $1.0$ are presented by the dark-yellow, 
purple and black solid lines, respectively. 
The dashed and dotted lines with different colors represent 
the different radiation components for the model of $\delta = 1.0$.
The observational data of the radio (magenta for \citet{Nayana2017}; green for \citet{Tian2008}), X-rays
\citep{Doroshenko2017}, TeV $\gamma$-rays \citep{Abramowski2011},
and GeV $\gamma$-rays presented in this work are shown.}
\label{fig:multi-sed}
\end{figure*}

For the leptonic model, the background radiation field considered includes
the cosmic microwave background (CMB), and an infrared (IR) radiation 
field with a temperature of 40 K and an energy density of 1 eV cm$^{-3}$ 
\citep{Abramowski2011}. The magnetic field strength is taken as a free
parameter, which is determined through fitting to the multi-wavelength
data. The derived model parameters are given in Table \ref{table:model}. 
The corresponding multi-wavelength SED of the model calculation is shown 
in the left panel of Figure \ref{fig:multi-sed}. 

The leptonic models with the three different values of $\delta$ can 
reproduce the muti-wavelength SED with little differences.
Compared with the results of \citet{Yang2014}, the spectral index of 
electrons $\alpha_e$ and cutoff energy $E_{c,e}$ are both slightly smaller in this work.
This may due to the updated radio, X-ray and GeV data we used in the model.
The magnetic field strength, $B \sim 28~\mu$G, 
is consistent with that given in \citet{Yang2014}. Such a magnetic field 
strength is slightly larger than that of several other SNRs which show 
similar GeV-TeV $\gamma$-ray spectra, e.g. RX J1713.7-3946 
\citep{Abdo2011,Yuan2011,Zeng2017}, RX J0852-4622 \citep[Vela Junior;][]
{Tanaka2011}, and RCW 86 \citep{Yuan2014}. These SNRs are believed to 
be a class of sources with leptonic origin of the $\gamma$-ray emission 
\citep{Yuan2012,Funk2015,Guo2017}. 



The cutoff of the spectrum may be due to the (synchrotron) cooling of
electrons. The synchrotron cooling time scale of HESS J1731-347 is 
estimated to be
\begin{equation}
t_{\rm syn} \approx 1800 \left(\frac{E_{c,e}}{9\, \mathrm{TeV}}\right)^{-1} \left(\frac{B}{28\,\mu\mathrm{G}}\right)^{-2}\, \mathrm{yr}. \nonumber
\end{equation}
This time scale is close to the minimum value of the age of HESS J1731-347
inferred with other methods 
\citep{Tian2008,Abramowski2011,Fukuda2014,Acero2015b}.

\citet{Nayana2017} reported an anti-correlation between the TeV $\gamma$-ray 
emission and radio brightness profile, and ascribed such an anti-correlation 
to the synchrotron cooling effect with a non-uniform magnetic field.
This result supports the leptonic scenario for the multi-wavelength
emission of HESS J1731-347.

The right panel of Figure \ref{fig:multi-sed} shows the multi-wavelength 
SED of the hadronic-leptonic hybrid model, in which the radio to X-ray
data is acounted for by the synchrotron emission of electrons, and the
GeV-TeV $\gamma$-ray emission is produced by the decay of neutral pions
from $pp$ collisions. The model parameters are also summarized in Table 
\ref{table:model}. 
For the hybrid models with different values of $\delta$, a hard spectral index
of protons with $\alpha_p \sim 1.7$ even $\alpha_p \sim 1.5$, is needed to explain the 
hard GeV $\gamma$-ray spectrum. However, such spectrum of protons is difficult to be produced 
in the conventional diffusive shock acceleration model of strong shocks. 
The total energy of protons above 1$~$GeV is estimated to be 
$W_p \sim 1.5 \times 10^{50} (n/1.0\,\mathrm{cm}^{-3})^{-1}\ 
(d/3.2\,\mathrm{kpc})^{2}\ \mathrm{erg}$, corresponding to $\sim$ 15\% 
particle acceleration efficiency for a typical total energy of 
$E_{\rm SN}\sim10^{51}$ erg released by a core-collapse supernova.
The total energy $W_p$ depends on the distance and ambient gas density 
of HESS J1731-347. Since there is no thermal X-ray emission observed, 
the gas density would be very low, (e.g., \citet{Abramowski2011} derived 
an upper limit of gas density of $\sim 0.01$ cm$^{-3}$ assuming an 
electron plasma temperature of 1 keV), and hence the corresponding 
$W_p$ would be much higher.
However, if HESS J1731-347 expands in an inhomogeneous environment with dense gas clumps, 
the hard $\gamma$-ray emission and the high energy budget can be solved \citep{Inoue2012,Gabici2014,Fukui2013}.

A spatial correlation between the TeV $\gamma$-ray shell and the 
interstellar protons at a distance of $\sim$ 5.2 kpc was reported in
\citet{Fukuda2014}. It was suggested that the hadronic process contributes 
a large fraction of the $\gamma$-ray emission of HESS J1731-347
\citep{Fukuda2014}. This is similar to the cases of RX J1713.7-3946 and 
RX J0852.0-4622 \citep{Fukui2012,Fukui2013}. However, no significant 
emission from dense molecular gas at such a distance was detected by 
\citet{Maxted2017}, which seems to be challenge to the hadronic scenario. 


\begin{table*}
\centering
\normalsize
\caption {Model parameters}
\begin{tabular}{cccccccccccccccc}
\hline \hline
Model& $\delta$& $\alpha_{p}$& $\alpha_{e}$&$B_{\rm SNR}$& $W_{e}$ & $E_{c,e}$ & $W_{p}$ & $E_{c,p}$  & $\chi^2$\\
     &           &    &             & ($\mu$G)     &($10^{47}$ erg) & (TeV)   &($10^{50}$ erg)& (TeV)    &\\ 
\hline
Leptonic  &$0.5$   & $-$  & $1.6$   & $29.0$   & $1.7$  & $1.2$ & $-$ & $-$ & $529.27/320$ \\
          & $0.6$  & $-$  & $1.7$   & $27.0$   & $2.0$  & $2.8$ & $-$ & $-$  &  $428.96/320$\\
          &$1.0$   & $-$  & $1.8$         & $28.0$      & $1.6$   & $8.9$  & $-$ & $-$ & $406.27/320$\\
\hline
Hybrid   & $0.5$  & $1.5$ & $1.6$        & $85.0$ & $0.35$ & $0.7$ & $1.4$ & $10.0$ & $472.04/317$ \\
         & $0.6$  & $1.5$ & $1.7$        & $87.0$ & $0.35$ & $1.5$ & $1.5$ & $15.0$ & $372.33/317$  \\
         & $1.0$  & $1.7$ & $1.8$        & $80.0$ & $0.3$ & $5.5$ & $1.5$ & $38.0$ & $377.31/317$\\
\hline \hline
\end{tabular}
\label{table:model}
\tablecomments{The total energy of relativistic particles, $W_{e,p}$, is
calculated for $E > 1$ GeV.}
\end{table*}

HESS J1729-345 is an unidentified TeV source near HESS J1731-347 
\citep{Abramowski2011}. Assuming that HESS J1731-347 locates at a
distance of $\sim$3.2 kpc, \citet{Cui2016} suggested that the TeV 
$\gamma$-ray emission of HESS J1729-345 possibly originates from the 
nearby molecular clouds illuminated by the CRs escaped from HESS J1731-347.
\citet{Capasso2016} reported a good spatial coincidence between the TeV 
$\gamma$-ray image in the bridge region and the dense gas at a distance 
of 3.2 kpc, which further supports the scenario of \cite{Cui2016}. 
\citet{Nayana2017} detected possible radio counterparts of HESS 
J1729-345 at 843 MHz and 1.4 GHz. 
However, the multi-wavelength data of HESS J1729-345 is still lack.
Future multi-wavelength observations are need to explore its nature.

\section{Conclusion}
In this paper, we report the GeV $\gamma$-ray emission from the direction
of HESS J1731-347 at a significance level of $\sim 4.7 \sigma$, with nine 
years of Pass 8 data recorded by the {\it Fermi}-LAT. The spatial morphology 
of HESS J1731-347 is found to be slightly extended in the GeV band.
The GeV spectrum can be described by a hard PL for with an index of 
$\Gamma = 1.77\pm0.14$. 

The $\gamma$-ray characteristics of HESS J1731-347 is similar with several
shell-type SNRs, including RX J1713.7-3946, RX J0852-4622, RCW 86, and
SN 1006. A pure leptonic model can account for the wide-band SED of HESS 
J1731-347. If the hadronic process is adopted to explain the $\gamma$-ray
emission, a very hard ($\sim1.6$) proton spectrum is required. In addition,
the energy budget of CR protons may be a problem, given a potentially low
gas density environment implied by the lack of thermal X-ray emission.

We also search for GeV $\gamma$-ray emission from the nearby source
HESS J1729-345. No significant excess is detected in its direction, and
the upper limits are given. More multi-wavelength observations are
necessary to address its emission mechanism, and test the proposed
scenario of the interaction between CRs escaped from HESS J1731-347
and the molecular clouds.

\section*{Acknowledgments}
We thank V. Doroshenko for providing the new X-ray data. 
This work is supported by National Key Program for Research and 
Development (2016YFA0400200), the National Natural Science 
Foundation of China (Nos. 11433009, 11525313, 11722328, 11703093), 
Natural Science Foundation of Jiangsu Province of China (No. BK20141444), 
and the 100 Talents program of Chinese Academy of Sciences.


\begin{thebibliography}{}

\bibitem[Abdo et al. (2010a)]{Abdo2010a}Abdo, A. A., Ackermann, M., Ajello, M., et al. 2010a, ApJL, 710, L92


\bibitem[Abdo et al. (2010b)]{Abdo2010b}Abdo, A. A., Ackermann, M., Ajello, M., et al. 2010b, Sci, 327, 1103

\bibitem[Abdo et al. (2011)]{Abdo2011}Abdo, A. A., Ackermann, M., Ajello, M., et al. 2011, ApJ, 734, 28

\bibitem[Abramowski et al. (2011)]{Abramowski2011}Abramowski, A., Acero, F., Aharonian, F., et al. 2011, A\&A, 531, A81

\bibitem[Abramowski et al. (2016)]{Abramowski2016}Abramowski, A., Aharonian, F., Ait Benkhali, F. et al. arXiv:1601.04461

\bibitem[Acciari et al. (2009)]{Acciari2009}Acciari, V. A., Aliu, E., Arlen, T., et al. 2009, ApJL, 698, L133


\bibitem[Acero et al. (2015a)]{Acero2015a}Acero, F., Ackermann, M., Ajello, M., et al. 2015, ApJS, 218, 23

\bibitem[Acero et al. (2010)]{Acero2010}Acero, F., Aharonian, F., Akhperjanian, A. G., et al. 2010, A\&A, 516, A62

\bibitem[Acero et al. (2015b)]{Acero2015b}Acero, F., Lemoine-Goumard, M., Renaud, M., et al. 2015, A\&A, 580, A74

\bibitem[Ackermann et al. (2013)]{Ackermann2013}Ackermann, M., Ajello, M., Allafort, A., et al. 2013, Sci, 339, 807

\bibitem[Aharonian et al. (2004)]{Aharonian2004}Aharonian, F., Akhperjanian, A. G., Aye, K.-M., et al. 2004, Natur, 432, 75

\bibitem[Aharonian et al. (2005)]{Aharonian2005}Aharonian, F., Akhperjanian, A. G., Bazer-Bachi, A. R., et al. 2005, A\&A, 437, L7

\bibitem[Aharonian et al. (2006)]{Aharonian2006}Aharonian, F., Akhperjanian, A. G., Bazer-Bachi, A. R., et al. 2006, A\&A, 449, 223

\bibitem[Aharonian et al. (2007a)]{Aharonian2007a}Aharonian, F., Akhperjanian, A. G., Bazer-Bachi, A. R., et al. 2007a, A\&A, 464, 235

\bibitem[Aharonian et al. (2007b)]{Aharonian2007b}Aharonian, F., Akhperjanian, A. G., Bazer-Bachi, A. R., et al. 2007b, ApJ, 661, 236

\bibitem[Aharonian et al. (2008a)]{Aharonian2008a}Aharonian, F., Akhperjanian, A. G., de Alemida, U. B., et al. 2008a, A\&A, 486, 829

\bibitem[Aharonian et al. (2008b)]{Aharonian2008b}Aharonian, F., Akhperjanian, A. G., de Almeida, U. B., et al. 2008b, A\&A, 477, 353

\bibitem[Aharonian et al. (2009)]{Aharonian2009} Aharonian, F., Akhperjanian, A. G., de Almeida, U. B., et al. 2009, ApJ, 692, 1500

\bibitem[Albert et al. (2007a)]{Albert2007a}Albert, J., Aliu, E., Anderhub, H., et al. 2007a, A\&A, 474, 937

\bibitem[Albert et al. (2007b)]{Albert2007b}Albert, J., Aliu, E., Anderhub, H., et al. 2007b, ApJL, 664, L87

\bibitem[Araya (2017)]{Araya2017}Araya, M, 2017, ApJ, 843, 12

\bibitem[Atwood et al. (2009)]{Atwood2009}Atwood, W. B., Abdo, A. A., Ackermann, M., et al. 2009, ApJ, 697, 1071

\bibitem[Bamba et al. (2012)]{Bamba2012}Bamba, A., P{\"u}hlhofer, G., Acero, F., et al. 2012, ApJ, 756, 149

\bibitem[Capasso et al. (2016)]{Capasso2016}Capasso, M., Condon, B., Coffaro, M., et al. 2016, arXiv:1612.00258

\bibitem[Condon et al. (2017)]{Condon2017}Condon, B., Lemoine-Goumard, M.,  Acero, F. \& Katagiri, H. 2017, arXiv:1711.05499

\bibitem[Cui et al.(2016)]{Cui2016} Cui, Y., P{\"u}hlhofer, G., \& Santangelo, A.\ 2016, A\&A, 591, A68 

\bibitem[Doroshenko et al. (2017)]{Doroshenko2017}Doroshenko, V., P{\"u}hlhofer, G., Bamba, A., et al. 2017, arXiv:1708.04110

\bibitem[Fukuda et al. (2014)]{Fukuda2014}Fukuda, T., Yoshiike, S., Sano, H., et al. 2014, ApJ, 788, 94

\bibitem[Fukui et al. (2012)]{Fukui2012}Fukui, Y., Sano, H., Sato, J., et al. 2012, ApJ, 746, 82

\bibitem[Fukui (2013)]{Fukui2013}Fukui, Y. 2013, in Astrophysics and Space Science Proc., Vol. 34, Cosmic
Rays in Star-Forming Environments, ed. D. F. Torres \& O. Reimer (Berlin: Springer), 249

%
\bibitem[Funk (2015)]{Funk2015}Funk, S. 2015, Annual Review of Nuclear and Particle Science, 65, 245

\bibitem[Gottschall et al. (2016)]{Gottschall2016}Gottschall, D., Capasso, M., Deil, C., et al. 2016, arXiv:1612.00261

\bibitem[Gabici \& Aharonian (2014)]{Gabici2014}Gabici, S. \& Aharonian, F. A. 2014, MNRAS, 445, L70

%
\bibitem[Guo et al. (2017)]{Guo2017} Guo, X.-L., Xin, Y.-L., Liao, N.-H., et al. 2017, ApJ, 835, 42

\bibitem[Halpern \& Gotthelf (2010)]{Halpern2010}Halpern, J. P., \& Gotthelf, E. V. 2010, ApJ, 710, 941

\bibitem[Hewitt et al. (2012)]{Hewitt2012}Hewitt, J. W., Grondin, M.-H., Lemoine-Goumard, M., et al. 2012, ApJ, 759, 89

\bibitem[Inoue et al. (2012)]{Inoue2012}Inoue, T., Yamazaki, R., Inutsuka, S., \& Fukui, Y. 2012, ApJ, 744, 71


\bibitem[Koyama et al. (1995)]{Koyama1995} Koyama, K., Petre, R., Gotthelf, E. V., et al. 1995, Nature, 378, 255

\bibitem[Li \& Chen (2010)]{Li2010} Li, H. \& Chen, Y. 2010, MNRAS, 409, L35

\bibitem[Li \& Chen (2012)]{Li2012} Li, H. \& Chen, Y. 2012, MNRAS, 421, 935

\bibitem[Maxted et al. (2017)]{Maxted2017}Maxted, N., Burton, M., Braiding, C., et al. 2017, arXiv:1710.06101

\bibitem[Nayana et al. (2017)]{Nayana2017} Nayana, A. J., Chandra, P., Roy, S., et al. 2017, 
MNRAS, 467, 155

\bibitem[Rieger et al. (2013)]{Rieger2013} Rieger, F. M., de O{\~n}a-Wilhelmi, E., \& Aharonian, F. A., 2013, Front. Phys., 8, 714


\bibitem[Tanaka et al. (2011)]{Tanaka2011}Tanaka, T., Allafort, A., Ballet, J., et al. 2011, ApJL, 740, L51

\bibitem[Tian et al. (2008)]{Tian2008} Tian, W.-W., Leahy, D. A., Haverkorn, M., \& Jiang, B. 2008, ApJL, 679, L85

\bibitem[Tian et al. (2010)]{Tian2010}Tian, W.-W., Li, Z., Leahy, D. A., et al. 2010, ApJ, 712, 790



\bibitem[Xin et al. (2016)]{Xin2016}Xin, Y.-L., Liang, Y.-F., Li, X., et al. 2016, ApJ, 817, 64

\bibitem[Xin et al. (2017)]{Xin2017} Xin, Y.-L., Guo, X.-L., Liao, N.-H., et al.\ 2017, ApJ, 843, 90 

\bibitem[Xing et al. (2016)]{Xing2016}Xing, Y., Wang, Z., Zhang, X., \& Chen, Y. 2016, ApJ, 823, 44

\bibitem[Yang et al. (2014)]{Yang2014}Yang, R.-Z., Zhang, X., Yuan, Q., \& Liu, S.-M. 2014, A\&A, 567, A23

\bibitem[Yuan et al. (2014)]{Yuan2014}Yuan, Q., Huang, X.-Y., Liu, S.-M., \& Zhang, B. 2014, 
ApJL, 785, L22

\bibitem[Yuan et al. (2012)]{Yuan2012}Yuan, Q., Liu, S.-M., \& Bi, X.-J. 2012, 
ApJ, 761, 133

\bibitem[Yuan et al. (2011)]{Yuan2011} Yuan, Q., Liu, S. M., Fan, Z. H., Bi, X. J., \& Fryer, C. L. 2011, ApJ, 735, 120

\bibitem[Zeng et al (2017)]{Zeng2017}Zeng, H.-D., Xin, Y.-L., Liu, S.-M., et al. 2017, ApJ, 834, 153

\end{thebibliography}
\end{document}